# Two-dimensional exciton-polariton - light guiding by transition metal dichalcogenide monolayers

Jacob B. Khurgin

Johns Hopkins University, Baltimore MD 21218

**Abstract:**

**A monolayer of transition metal dichalcogenide (TMDC) is shown to be capable of supporting a guided optical mode below the exciton resonance – a two-dimensional exciton polariton. This visible or near IR mode is confined roughly with a micrometer from the monolayer and has propagation length exceeding 100 micrometers. The light guiding ability makes TMDC monolayers more versatile and potentially attractive photonic platform.**

A monolayer-thick two dimensional material have become objects of intense studies due to their potential applications in electronics and optics [1, 2]. Despite their thickness measured in angstroms, 2D materials can support substantial electric currents, as well as absorb and emit measurable amount of light. Among the optical properties, ability to support a guided mode is among the most fundamental and consequential ones. Graphene, the original 2D material is known to support a two-dimensional surface-plasmon polariton, largely confined near the graphene sheet and having large propagation constant and slow group velocity [3]. The graphene plasmonics is a thriving field where numerous potential applications are being investigated. While plasmons are transverse-magnetic (TM) waves existing in media with negative permittivity, it had been shown in [4] that in the narrow interval where permittivity of graphene turns positive, yet below absorption edge a weakly confined transverse electric (TE) can also exist. However, guided waves in graphene require high doping, and are restricted to a frequency region below the absorption edge occurring at twice Fermi energy. It follows that graphene photonics is practicable mostly in the range from THz to perhaps mid-infrared [5], and if one is to expand this range to more interesting visible or UV range, a monolayer material with preferably wide direct bandgap should be considered. In the last few years a family of such materials, transition metal dichalcogenides (TMDC) has been identified [6] and high quality monolayers of TMDC's, $MoS_2$ and $WSe_2$ being most prominent representatives, have been fabricated and explored, revealing strong excitonic features in the visible range [7,8]. Due to large effective mass and small dielectric constant of the surrounding material peak excitonic absorption in TMDC's reaches 15-20% per monolayer and the exciton binding energy is measured in hundreds of meV, making excitonic feature stable at elevated temperatures. Strong exciton-photon coupling had been observed in microcavities [9]. It is then reasonable to expect that a two-dimensional exciton can couple with a propagating photon and engender a two-dimensional exciton-polariton – a confined quasiparticle propagating in single atomic layer of TMDC. In this work we show the conditions under which guided-mode 2D exciton-polariton in TMDC monolayer exist and estimate its most relative characteristics – dispersion, confinement factor and propagation length.

The TMDC exciton polariton, shown in Fig.1 is a TE wave propagating along a TMD monolayer placed inside the dielectric cladding with index n, with the fields described as

$$E_y = E_0 e^{\mp qx} e^{i(\beta z - \omega t)}$$

$$H_x = -\frac{\beta}{\omega\mu_0} E_0 e^{\mp qx} e^{i(\beta z - \omega t)}, \qquad (1)$$

$$H_z = \pm \frac{iq}{\omega\mu_0} E_0 e^{\mp qx} e^{i(\beta z - \omega t)}$$

where the upper sign correspond to x>0 and lower one to x<0, and $\beta^2 - q^2 = n^2\omega^2/c^2$. The tangential electric field engenders surface polarization $P_y^{(s)} = \varepsilon_0 \chi^{(s)} E_y$ where $\chi^{(s)}(\omega) = \chi_r^{(s)}(\omega) + \chi_i^{(s)}(\omega)$ is a complex surface susceptibility (in units of length).

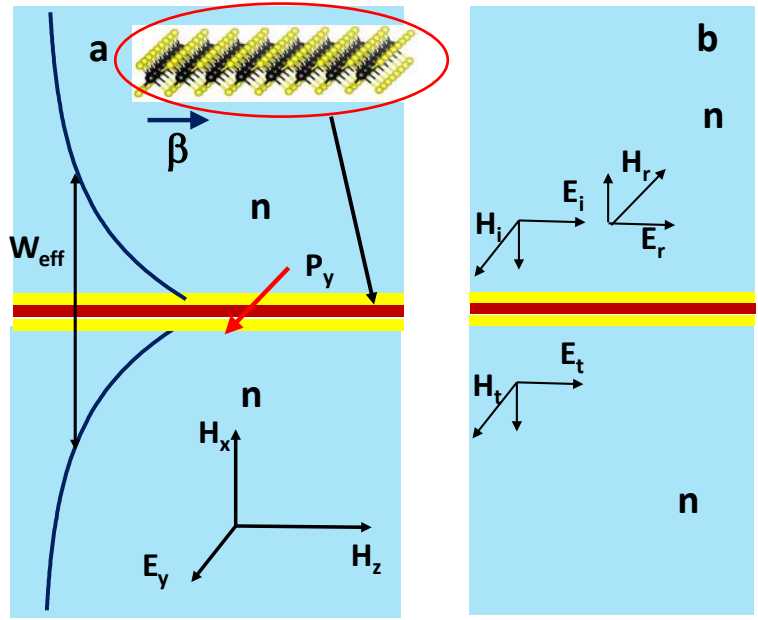

Fig. 1 a) Fields in Exciton Polariton supported by a TMDC monolayer b) Absorption and reflection by a TMDC monolayer

Tangential component of magnetic field is subject to the Maxwell equation $\partial H_z / \partial x = i\omega P_y$, and integrating it along the contour $\delta l$ encompassing TMDC monolayer yields a boundary condition

$$H_z(+0) - H_z(-0) = i\omega P_y^{(s)}, \qquad (2)$$

where $i\omega P_y^{(s)}$ is a surface density of the polarization current. Combining (2) with the third equation in (1) we obtain

$$\frac{2iq}{\omega\mu_0} E_0 = i\omega\varepsilon_0 \chi^{(s)} E_0. \qquad (3)$$

This allows us to find the complex decay constant $q(\omega) = \omega^2 \chi^{(s)} / 2c^2$ and define he dispersion relationship as $\beta(\omega) \approx k_d \left(1 + (1/2)(\chi^{(s)} \omega / 2cn)^2\right)$.

The t2D susceptibility in (3) is directly related absorption by the TMDC monolayer of the harmonic wave propagating in the direction normal to it (Fig.1b). Using the boundary condition for tangential electric and magnetic fields

$$\begin{aligned} E_i + E_r &= E_t \\ H_i - H_r &= H_t - i\omega \chi^{(s)} E_t \end{aligned} \quad (4)$$

Solving these equation one obtains the relation between the incident and transmitted amplitudes of electric field $E_i = E_t(1 - i\omega \chi^{(s)} / 2nc)$. Assuming that absorption is small, and keeping only the lowest order term one obtains $|E_t|^2 = |E_i|^2 - \alpha(\omega)|E_i|^2$, where the absorption coefficient is $\alpha(\omega) = \omega \chi_i^{(s)}(\omega) / nc$.

If we now approximate the excitonic absorption by a Lorentzian shape with peak absorption at exciton resonance $\alpha(\omega_0) = \alpha_{max}$ and FWHM linewidth $\gamma$ (Fig.2) the 2D susceptibility can be evaluated as

$$\chi^{(s)}(\omega) = \frac{\alpha_{max} cn}{\omega} \frac{\gamma/2}{\omega_0 - \omega - i\gamma/2} \quad (5)$$

The decay coefficient can now be found as

$$\frac{q(\omega)}{k_d} = \frac{\alpha_{max} \gamma}{4} \frac{1}{(\omega_0 - \omega) - i\gamma/2}, \quad (6)$$

where $k_d = n\omega/c$ is a wavevector of free wave in cladding material, and the dispersion relation becomes

$$\frac{\beta(\omega)}{k_d} \approx 1 + \frac{\alpha_{max}^2 \gamma^2}{32} \left(\frac{(\omega_0 - \omega) + i\gamma/2}{(\omega_0 - \omega)^2 + \gamma^2/4}\right)^2 \quad (7)$$

That allows us to determine the effective width of the polariton mode,

$$W_{eff}(\omega) = \frac{1}{\text{Re}(q)} = \frac{2\lambda_d}{\pi \alpha_{max} \gamma} \frac{(\omega_0 - \omega)^2 + \gamma^2/4}{\omega_0 - \omega} \approx \frac{2\lambda_d (\omega_0 - \omega)}{\pi \alpha_{max} \gamma} \quad (8)$$

where $\lambda_d$ is the wavelength in the cladding dielectric, and the propagation length

$$L_p = \frac{1}{2\,\text{Im}(\beta)} = \frac{8\lambda_d}{\pi \alpha_{max}^2 \gamma^3} \frac{\left[(\omega_0 - \omega)^2 + \gamma^2/4\right]^2}{\omega_0 - \omega} \approx \frac{8\lambda_d (\omega_0 - \omega)^3}{\pi \alpha_{max}^2 \gamma^3} \quad (9)$$

The rightmost expressions in (8) and (9) are the off-resonance approximations. The minimal effective width of polariton, occurring at, $\omega = \omega_0 - \gamma/2$ depends only on peak absorption, as $W_{min} = 2\lambda_d / \pi \alpha_{max}$ Interestingly, far from the resonance effective width depends only on the area under the exciton peak, $\alpha_{max} \gamma$, or, essentially on the oscillator strength of the exciton. A rough, order-of-magnitude estimate of the oscillator strength can be obtained assuming exciton contains the entire oscillator strength of the electron-hole states whose wave-vectors are within inverse Bohr radius $a_B^{-1}$, i.e. whose energies are within binding energy $E_B$, i.e. $\alpha_{max}\hbar\gamma \simeq \pi\alpha_0 E_B$, where $\alpha_0$ is a fine structure constant. We can then write for the off-resonant effective width, $W_{eff}(\Delta E) \approx 2\lambda_d \Delta E / \pi^2 \alpha_0 E_B$, where $\Delta E = \hbar(\omega_0 - \omega)$

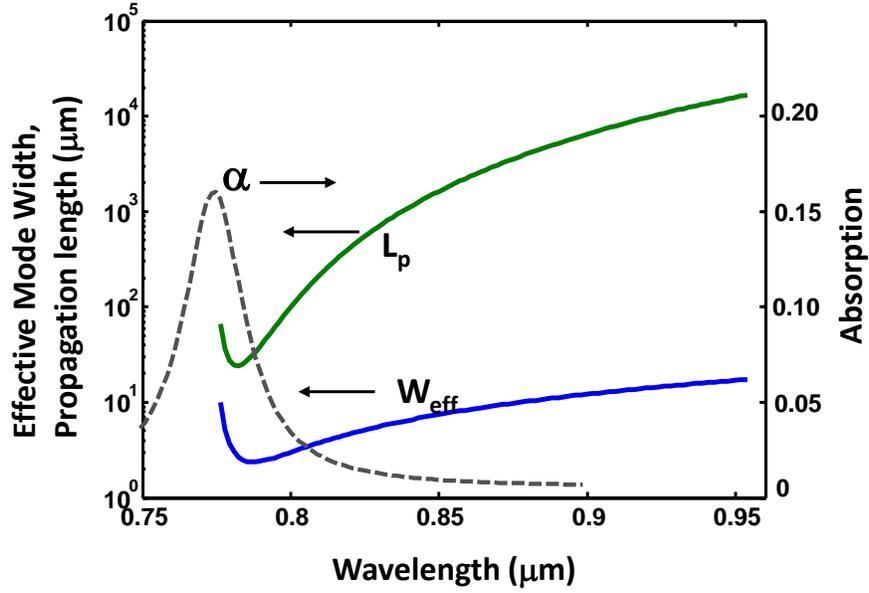

Fig. 2 Effective mode width $W_{eff}$, and propagation length $L_p$ of exciton-polariton in WSe$_2$ whose normal incidence absorption $\alpha$ is also shown

To provide an example we rely on the results obtained in Ref [8], where absorption by a monolayer of WSe$_2$ placed on the quartz substrate (n=1.41) had been studied. Maximum excitonic absorption of $\alpha_{max} = 0.15$ with FWHM linewidth of $\hbar\gamma = 50 meV$ was measured at the resonance energy $\hbar\omega_0 = 1.6 eV$ (A-exciton), and the binding energy was estimated to be $E_B = 0.37 eV$. Note that our rough estimate of exciton oscillator strength for this example is correct within 15%, even though the excitonic radius is only 8A indicating that Wannier-Mott exciton model is not really well applicable.

Using these values in (8) and (9), the effective width and propagation length have been determined and are shown in Fig.2. The minimum effective width occurs at the wavelength $\lambda = 788 nm$ and equals 2.3µm. The propagation length, however is rather short, 30 µm. But at a slightly longer wavelength $\lambda = 800 nm$ the propagation length exceeds 100 µm, while the effective mode size is only 3 µm. It should be noted that while effective mode size is larger than in a conventional "thick" waveguide, large mode size does have advantages. First of all, the wide mode is less sensible to the small surface perturbations, and, second, with wide mode coupling in and out of fiber gets easier.

It is also interesting to compare these results with a conventional "3D" dielectric waveguide. For that we estimate the maximum surface (sheet) susceptibility from (5) as $\chi^{(s)}_{r,max} = \alpha_{max}\lambda n / 4\pi \approx 13 nm$. If we considered a waveguide made, say from heavy glass with n$_g$=1.85 the thickness required to provide similar sheet susceptibility would be $t_g = 13 nm / (n_g^2 - n^2) \approx 9 nm$. Thus a monolayer of WSe$_2$ has the same confinement ability as a slab of a 3D material that is nearly 30 times thicker.

Another important observation is that confined mode exists only if the refractive indices on both sides of the monolayer are close to each other. If one retraces the derivation for the case where the indices on two sides differ by small amount $\Delta n$, then the confined mode exits only for as long as

$$\Delta n < \left|\frac{\omega}{c}\chi^{(s)}(\omega)\right|^2 / 2n = \frac{\alpha_{max}^2 n}{4}\frac{\gamma^2/2}{(\omega_0-\omega)^2+\gamma^2/4}, \qquad (10)$$

So for the detuning of HWFM the indices of refraction should differ by no more than 1% which is of course within realm of possible.

Our final observation is that if anything, in this simple model we have underestimated the light guiding performance of TMDC monolayers, because there is an additional contribution to the susceptibility due to split-off exciton B situated about 100 meV above the main exciton A [6,8]. Furthermore, the propagation away from resonance may be even longer than estimated here because away from the resonance absorption decays exponentially [10] (Urbach tail) rather than quadratically as in Lorentzian form.

In conclusion we have demonstrated theoretically that a single monolayer of transition metal dichalcogenides is capable of supporting confined exciton-polariton modes in the visible and near IR ranges with effective width of just a few micrometers and propagation length in excess of 100 micrometers. It is reasonable, in our view, to expect that these characteristics will be further improved once higher quality materials with narrower excitonic linewidth become available. Besides being scientifically notable, this impressive light guiding ability of TMDC monolayers enhances the chances that TMDC's will find practical niche in nanophotonics.